\documentclass{article}

\def\BibTeX{{\rm B\kern-.05em{\sc i\kern-.025em b}\kern-.08em
    T\kern-.1667em\lower.7ex\hbox{E}\kern-.125emX}}
\usepackage{PRIMEarxiv}

\usepackage[utf8]{inputenc} % allow utf-8 input
\usepackage[T1]{fontenc}    % use 8-bit T1 fonts
\usepackage{hyperref}       % hyperlinks
\usepackage{url}            % simple URL typesetting
\usepackage{booktabs}       % professional-quality tables
\usepackage{amsfonts}       % blackboard math symbols
\usepackage{nicefrac}       % compact symbols for 1/2, etc.
\usepackage{microtype}      % microtypography
\usepackage{lipsum}
\usepackage{fancyhdr}       % header
\usepackage{graphicx}       % graphics
\graphicspath{{media/}}     % organize your images and other figures under media/ folder
\usepackage{amsmath}
\usepackage{derivative}
\usepackage{amssymb}
\title{Low-Dose CT Denoising via Sinogram Inner-Structure Transformer
}

\author{Liutao~Yang, Zhongnian~Li, Rongjun~Ge, Junyong~Zhao, Haipeng~Si, and Daoqiang~Zhang
}
\begin{document}
\maketitle

%{Author \MakeLowercase{Liutao Yang\textit{et al.}}: Low-Dose CT Denoising via Sinogram Inner-Structure Transformer}

\begin{abstract}
Low-Dose Computed Tomography (LDCT) technique, which reduces the radiation harm to human bodies, is now attracting increasing interest in the medical imaging field. As the image quality is degraded by low dose radiation, LDCT exams require specialized reconstruction methods or denoising algorithms. However, most of the recent effective methods overlook the inner-structure of the original projection data (sinogram) which limits their denoising ability. 
The inner-structure of the sinogram represents special characteristics of the data in the sinogram domain. By maintaining this structure while denoising, the noise can be obviously restrained. 
Therefore, we propose an LDCT denoising network namely Sinogram Inner-Structure Transformer (SIST) to reduce the noise by utilizing the inner-structure in the sinogram domain. Specifically, we study the CT imaging mechanism and statistical characteristics of sinogram to design the sinogram inner-structure loss including the global and local inner-structure for restoring high-quality CT images. 
Besides, we propose a sinogram transformer module to better extract sinogram features. The transformer architecture using a self-attention mechanism can exploit interrelations between projections of different view angles, which achieves an outstanding performance in sinogram denoising.  Furthermore, in order to improve the performance in the image domain, we propose the image reconstruction module to complementarily denoise both in the sinogram and image domain. 
\end{abstract}

\section{Introduction}
\label{sec:introduction}

%\IEEEPARstart{T}{his} 
Computed Tomography (CT) is an important medical imaging modality for medical diagnosis. However, exposure of patients to the X-ray radiation of CT examinations would increase the risk of cancer. Therefore, the Low-does Computed Tomography (LDCT) technique is in urgent need in the clinic. The most common way to reduce the radiation dose in CT scans is to reduce the X-ray tube current (or voltage). As a consequence, images that are reconstructed from this current (or voltage) always suffer from severe noise.

\begin{figure*}
	\begin{center}
		%\fbox{\rule{0pt}{2in} \rule{1\linewidth}{0pt}}
		\includegraphics[width=1\linewidth]{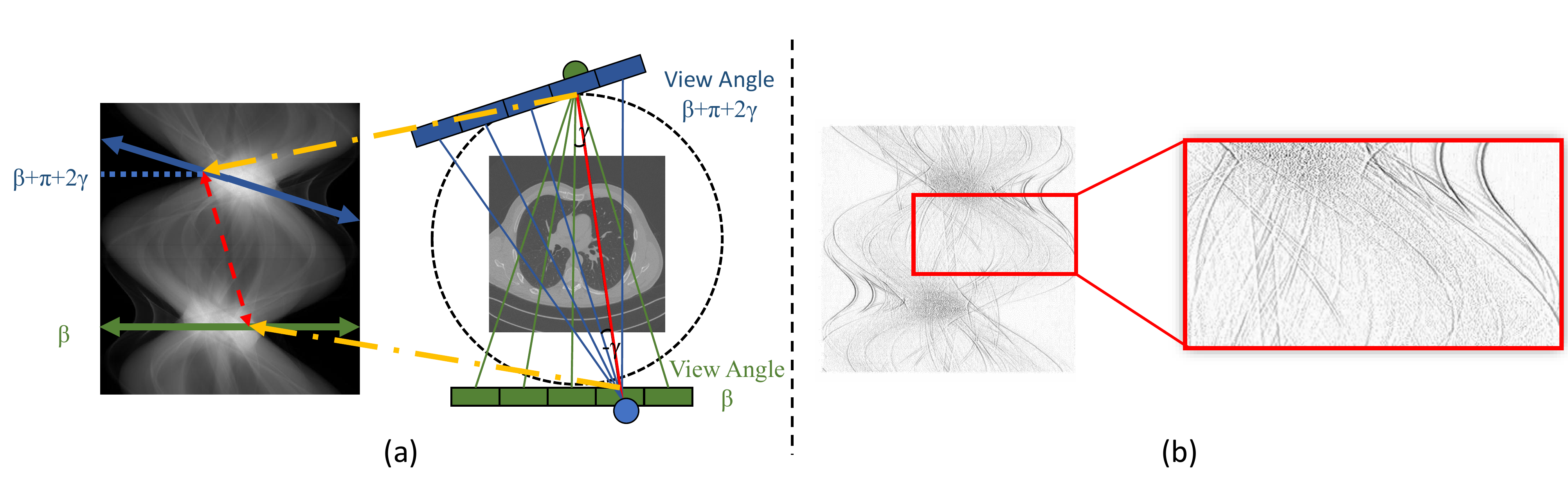}
		
	\end{center}
	\caption{Illustration of the sinogram inner-structure that we use for sinogram denoising. (a) shows conjugate projection pairs in the sinogram. Each row of the sinogram represents the projections under one view angle. In the figure, conjugate projections under view angle $\beta$ (green) are corresponding to the blue line. The yellow arrows show one example of conjugate projection pair. (b) is the horizontal second-order derivatives of this sinogram. We can see the second-order derivatives are of high sparsity. By maintaining these inner-structures while sinogram denoising, we can significantly improve the image quality.   }
	\label{motivation}
\end{figure*}

In order to effectively improve the imaging quality of LDCT, lots of methods are developed in the past decades. They can be roughly divided into two categories: 1) sinogram domain reconstruction and 2) image domain post-processing. Sinogram domain methods focus on the original projection data. They either use filters \cite{balda2012ray, manduca2009projection} on the projection data to smooth the sinogram or use iterative reconstruction \cite{chen2020airnet,xia2019spectral,wu2017iterative,stayman2012model,geyer2015state,chen2013limited,zhang2016statistical} based on priors. However, filtration methods always result in spatial resolution loss in the reconstructed image.
Iterative reconstruction methods are time-consuming. Since the deep learning methods \cite{chen2019synergistic,cao2020auto,miao2018dilated,li2019segan} achieve great progress in medical image segmentation, reconstruction, and analysis, image domain post-processing \cite{chen2011improving,chen2012thoracic,ha2015low,zhang2016artifact,chen2017low,shan2019competitive} methods of LDCT denoising attain a state-of-the-art performance by utilizing the deep learning framework. 
As pure image domain methods overlook the origin projection data, several methods \cite{yin19, zhang2021clear} add the sinogram domain information into the deep learning denoising network to further improve the performance. However, these methods, without consideration of the inner-structure of original projection data, have limitations on the denoising ability.
Due to the physical mechanism of CT imaging, the sinogram contains a special inner-structure compared to the natural image. For example, as shown in Fig.\ref{motivation} (a), conjugate sampling pairs during CT scans receive X-rays with the same path so that corresponding points in the sinogram should obtain the same value. For each point of the sinogram, there is a corresponding conjugate point resulting in a conjugate structure over the sinogram domain. Considering that noise appears randomly (e.g, Poisson Distribution) in the whole imaging procedure, the conjugate pairs obtaining different noise would evidently break this structure. By maintaining this sinogram inner-structure, we can effectively restrain the noise and improve the image quality. Therefore, exploring the inner-structure of sinogram is of great importance for sinogram domain denoising. 
%for improving the denoising performance in LDCT

In this paper, we study the CT imaging mechanism and statistical characteristics to propose the Sinogram Inner-Structure Loss (SISL). As we illustrated above, the sinogram inner-structure represents the typical data characteristics in the sinogram domain. And the noise, which appears randomly while imaging, would break this structure. Designing a loss function based on sinogram inner-structure for network training in the sinogram domain can effectively maintain this structure and restrain noise. The inner-structure loss is designed both at the global and local levels. The global inner-structure utilizes conjugate sampling pairs in CT scans. As Fig.\ref{motivation} (a) shows, this relation constrains conjugate pairs in the sinogram to obtain the same value. Since the correlation exists in the whole sinogram, it can help maintain the global inner-structure. The local inner-structure considers the sparsity of second-order derivatives in the sinogram domain. As illustrated in \cite{xie2017robust}, the sinogram can be described as a piecewise linear configuration. So, we can obtain that second-order derivatives of the sinogram will be very sparse. Fig.\ref{motivation} (b) shows horizontal second-order derivatives of the sinogram. This feature is calculated between adjacent points in the sinogram which can help maintain the local inner-structure. 

%The global inner-structure utilizes conjugate sampling pairs in CT scans to constraint the corresponding points to obtain the same value. Since the correlation exists in the whole sinogram, it can enhance the global structure. The local inner-structure considers the second-order derivatives sparsity of the sinogram. As illustrated in \cite{xie2017robust}, the sinogram can be described as a piecewise linear configuration. So, we can obtain that the second-order derivatives of sinogram will be very sparse. Fig \ref{motivation}(b) shows the horizontal second-order derivatives of the sinogram. This feature indicates the correlation between the adjacent points in the sinogram which represents the local structure of sinogram. 

For network design, most exit methods use a CNN-based backbone in both the image and sinogram domain. However, we argue that the transformer architecture is more applicable in the sinogram domain. The transformer shows outstanding performance in NLP field by utilizing the multi-head self-attention to extract interrelation between sequence data. According to the CT imaging mechanism, each row of the sinogram is the projection at a certain view angle. By regarding the sinogram as sequences of projection, the self-attention mechanism in the transformer can extract relations between projections under different view angles, which is hard to achieve in a CNN architecture. %So that it can achieve better denoising performance than CNN-based networks.
Furthermore, since pure sinogram domain denoising can lead to artifacts in reconstructed images, we use an image reconstruction module to transfer the sinogram noise into the image domain and apply image domain denoising in one unified network. Thus, the image domain loss can be back-propagated into the sinogram domain for complementary optimization.
%Furthermore, in order to supervise the sinogram with the image domain ground truth, we use an image reconstruction module to transfer the sinogram noise into image domain and apply image denoising once more in one end-to-end network.
% According to the CT imaging mechanism, each row of sinogram is the projection for the whole image at a certain view angle, which indicates the sinogram is a sequence of projections. The transformer shows the outstanding performance in NLP fields. Chen \textit{et al}. \cite{chen2021pre} developed a image denoising transformer which achieves the SOTA performance. The 

% For denoising network design, we notice that convolutional operation in CNNs is no longer suitable for sinogram structure. According to the CT imaging mechanism, each row of sinogram is the projection for the whole image at a certain view angle, which indicates the sinogram is a sequence of projections. The typical CNN usually extract features using square kernels which breaks the sequence structure and degrade the performance. 
%Thus, we propose a LDCT denoising network namely Sinogram Inner-Structure Transformer (SIST). This network use a transformer-based architecture taking projections in the same view angle (one row in sinogram) as the input to better extract feature for sinogram. 

Contributions of this paper can be summarized as follows:

\begin{itemize}
\item We propose the global inner-structure loss which utilizes conjugate sampling pairs in CT scans and local inner-structure loss which considers the second-order sparsity of sinograms. These losses can help to better use the sinogram information and improve the denoising performance.

\item We propose a Sinogram Inner-Structure Transformer (SIST) for LDCT denoising. This network, taking each view angle of sinogram as input, can more effectively extract the structure feature than CNN-based networks.

\item We propose an image reconstruction module to transfer the sinogram noise into the image domain and apply image domain denoising in one unified network. Thus, the image domain loss can be back-propagated into the sinogram domain for complementary optimization.
\end{itemize}

The rest of the paper is organized as follows: In Related Works, we briefly introduce the LDCT reconstruction/denoising methods and transformers used in computer vision. Then, in the Proposed Method, we give the detail of our proposed Sinogram Inner-Structure Transformer. In Experiments, we compare the proposed method with state of the art and conduct several ablation studies. Finally, we give a conclusion in the last section. 

\section{Related Works}
\subsection{Low Dose CT Reconstruction and Denoising}
Due to the importance of LDCT, lots of model-based image reconstruction methods have been proposed during the past decades. These methods, which are based on priors of CT images, system model, or measurement statistical model, iteratively find the optimal image from the projection data. Ma \emph{et al.} \cite{ma2011low} proposed the previous normal-dose scan induced nonlocal means (ndiNLM) method to utilize the normal-dose image to enable low dose CT image reconstruction. Dose reduction using prior image constrained compressed sensing (DRPICCS) \cite{lauzier2013characterization} is proposed to reduce image noise using compressed sensing. Recently proposed PWLS-ULTRA \cite{zheng2018pwls} exploits loss based on an efficient Union of Learned
TRAnsforms which is pre-learned from numerous image patches extracted from a dataset of CT images. By using the learning-based method in iterative reconstruction, this method can achieve a good performance.

Different from the model-based image reconstruction methods focus on the reconstruction phase, deep learning based CT image denoising tries to remove the artifact and noise in reconstructed images. Inspired by the SRCNN \cite{dong2014learning}, Chen \emph{et al.} \cite{chen2017low} proposed a residual encoder-decoder CNN (REDCNN) for LDCT. With the development of GAN, Wolterink \textit{et al.} \cite{wolterink2017generative} propose a method to train a generative CNN jointly with an adversarial CNN to estimate NDCT images from LDCT images and hence reduce noise. Then Shan \emph{et al.} \cite{shan2019competitive} proposed a modularized adaptive processing neural network (MAP-NN), which performs an end-to-process mapping with a modularized neural network to optimize the denoising depth in a task-specific fashion. Yin \emph{et al.} \cite{yin19} proposed a domain progressive 3D residual convolution network (DP-ResNet) to denoising both in sinogram and image domain. This method train sinogram denoising CNN and image denoising CNN separately. Then they reconstruct the denoised sinogram into images and feed into the image domain CNN for further denoising. However, all those deep learning methods either only consider the image domain information or overlook the inner-structure of original projection data. Thus, our proposed method which uses the sinogram inner-structure to assist the LDCT denoising can considerably improve the performance.
\begin{figure*}[t]
	\begin{center}
		%\fbox{\rule{0pt}{2in} \rule{1\linewidth}{0pt}}
		\includegraphics[width=1\linewidth]{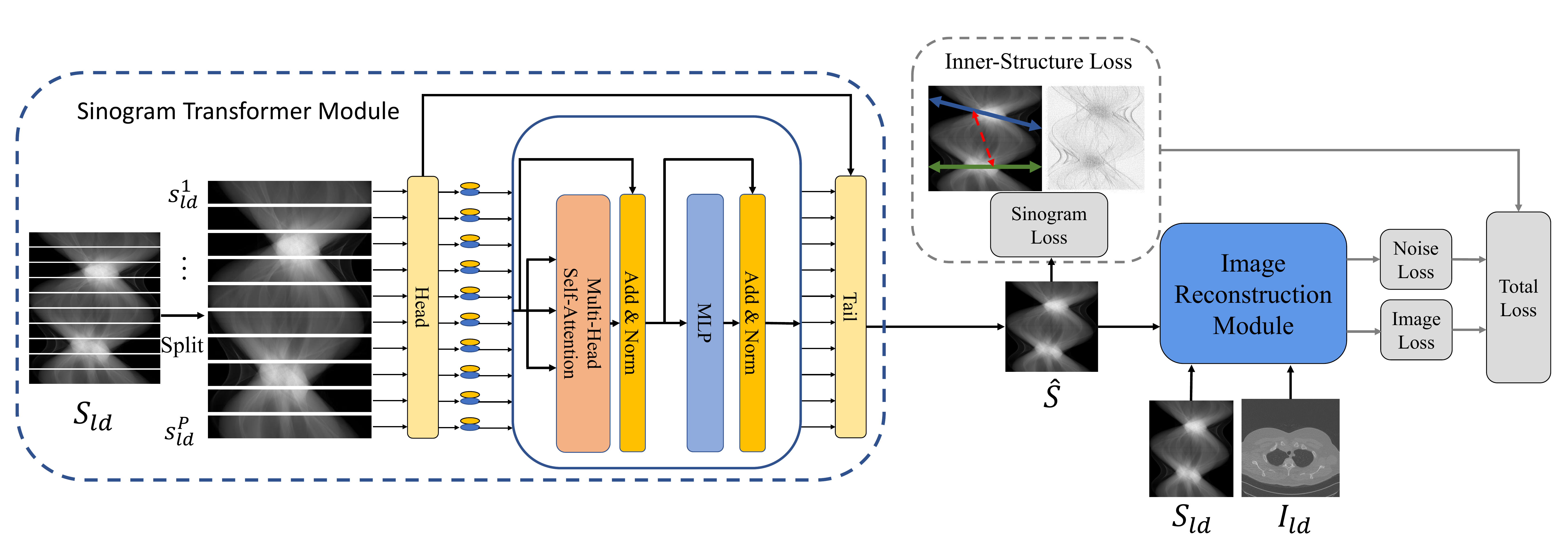}
		
	\end{center}
	\caption{The overall architecture of our proposed method. In this figure, $S_{ld}$ and $I_{ld}$ are the LDCT sinogram and image, $\hat{S}$ and $\hat{I}$ are output sinogram and image, $\hat{S}_{noise}$ and $\hat{I}_{noise}$ are the sinogram noise and image noise. $S_{ld}$ is first input into the sinogram transformer module for sinogram domain denoising. In this part, we use the proposed inner-structure loss to improve the denoising performance. Then the denoised sinogram $\hat{S}$ is input into the image reconstruction module for image domain denoising. Details of the image reconstruction module are shown in Fig.\ref{imre}. }
	\label{fig2}
\end{figure*}	

\subsection{Transformers in Computer Vision}
The transformer is first proposed \cite{vaswani2017attention} for natural language processing. Due to the competitive performance in many tasks, it soon became a very important architecture used in Natural Language Processing (NLP). For example, GPT \cite{radford2018improving,radford2019language} pre-trains in an auto-regressive way that predicts the next word in huge text datasets. BERT \cite{devlin2018bert} utilizes the transformer to predict a masking word based on context.

The success of transformer in NLP encourage researchers to attempt its applications in computer vision task. Dosovitskiy \emph{et al.} proposed the Vison Transformer (ViT) \cite{dosovitskiy2020image} to directly split the image into patches and provide the sequence of linear embeddings. This sample but effective architecture attains excellent results in image classification. Wu \emph{et al.} \cite{wu2020visual} represent images as semantic visual tokens and run transformers to densely model token relationships. This method achieves impressive results both in image classification and semantic segmentation. Hassani \emph{et al.} \cite{hassani2021escaping} propose the Compact Transformer using convolutional tokenization in transformer head and achieves state-of-the-art performance in small datasets classification. Pre-trained image processing transformer (PIT)\cite{chen2021pre} introduces the transformer into the low-level computer vision task (e.g., denoising, super-resolution, and deraining). With convolutional tokenization and pretrained on large datasets, this method also achieves the state of the art.

In all these works above, we observe that the key point for vision transformer is the tokenization method. In another word, how to transfer the 2D structure image into the 1D structure as input can largely impact the performance of vision transformers. As the sinogram is essentially 1D projection data, the transformer can better extract the information than a CNN architecture. So in this paper, we design the sinogram transformer for LDCT denoising.

\section{Proposed Method}
In this section, we introduce each part of the proposed Sinogram Inner-Structure Transformer (SIST).
 We first introduce the sinogram transformer module for sinogram domain denoising. Then we give the details of the Sinogram Inner-Structure Loss (SISL). Besides, we introduce the image reconstruction module which converts sinogram noise into image domain for complementarily denoising both in sinogram and image domain. Finally, we give the total loss function used for training. The overall architecture is shown in Fig.\ref{fig2}.

%In SIST, the LDCT sinogram is split into sequence data as the input of the sinogram transformer module for sinogram domain denoising. In sinogram domain, besides the sinogram loss supervised with groundtruth,  we add both global and local inner-structure loss to maintain the structure of sinogram while training. At last, the sinogram noise is converted into image domain for computing the noise transfer and image denoising loss. Thus, the loss in image domain can be back propagated to guide the sinogram training. 

\subsection{Preliminaries}
The LDCT denoising task in this paper is to recover the NDCT sinogram $S_{nd} \in \mathbb{R}^{P \times D }$ and image $I_{nd}\in \mathbb{R}^{W \times H } $ from the LDCT sinogram $S_{ld} \in \mathbb{R}^{P \times D }$ and image $I_{ld}\in \mathbb{R}^{W \times H } $, where $P$ is the number of projection views, $D$ is the number of detectors, $W$ and $H$ are the width and height of images. $s^{i}_{ld} \in \mathbb{R}^{D}, i=\{1,2,...,P\}$ is projections in $i_{th}$  view angle of $S_{ld}$. Note that $I_{nd}$ and $I_{ld}$ can be reconstructed from the sinogram $I_{nd}$ and $I_{ld}$ using reconstruction algorithm such as Filtered Back-Projection (FBP). $\hat{S}$, $\hat{I}$ and $\hat{I}_{noise}$ are the generated sinogram, image and image noise respectively.

\subsection{Sinogram Transformer Module}
The transformer architecture can more effectively extract the sinogram feature from different view angles than CNN-based networks due to the multi-head self-attention mechanism. So, we propose the sinogram transformer module for sinogram domain denoising.
Same as the typical transformer design, the sinogram transformer module consists of three parts: head, transformer encoder, and tail. The detail of each part are shown as follows:  
\subsubsection{Head.}
%Typical vision transformer split the image or convolutional feature maps into patches. This design is for the extraction of local information in the nature images. However, this definition is no longer suitable in the sinogram.
%we redesign it in the head. Each row of sinogram is consist of projections in the same view angle, so
Since tokenization is the key component to the performance, we split each row of the sinogram as a set of sequence data. Then we use an MLP to embed the sinogram into the input:
\begin{eqnarray}
\label{eq1}
    H=ReLU(LN([s^{1}_{ld},s^{2}_{ld},...,s^{P}_{ld}]))_{\times n}
\end{eqnarray}
where LN is the linear layer convert sinogram $s^{i}_{ld}$ with dimension $D$ into embeddings dimension $D'$ . ReLU is the activate function. $ H \in \mathbb{R}^{H \times D' }$ is the extracted feature as the input of transformer. The MLP consists of this block repeated for $n$ times.
 
\subsubsection{Transformer Encoder.}
In this part, we follow the original Vision Transformer design. Details can refer to \cite{vaswani2017attention}.
We add the learnable position encodings $e_{i}\in \mathbb{R}^{D'} $ to each projection of the transformer input $h_{i}$ to maintain the position information \cite{carion2020end,dosovitskiy2020image} . Then $e_{i}+h_{i}$ is input into the transformer encoder directly. The transformer encoder consists of Multi-head Self-Attention (MSA) and MLP. The entire structure is shown as follows:

\begin{eqnarray}
\begin{aligned}
&Z_{0}=[e_{1}+ h_{1};e_{2}+ h_{2};...;e_{P}+ h_{P}],\\
&Q_{i}=K_{i}=V_{i}=LN(Z_{i-1})\\
&Z^{'}_{i}=MSA(Q,K,V)+Z_{i-1},,  &i=1,...,l\\
&Z_{i}=MLP(Z^{'}_{i})+Z^{'}_{i}  \\
\end{aligned}
\end{eqnarray}

where $l$ is the number of layers in the transformer encoder. $LN$ is the linear layer.

\subsubsection{Tail.}
In the tail, We add a skip connection from the head to maintain the low-level feature. We use an MLP to recover the output of transformer encoder from dimension $D'$ into $D$:
\begin{eqnarray}
S^{'}=ReLU(LN(Z_{l}+H))_{\times n}
\end{eqnarray}
Then the output $S^{'}$ is restored to the shape $\mathbb{R}^{P \times D }$.  Finally, residual blocks are used for the final refinement of the sinogram:
\begin{eqnarray}
\hat{S}=Residual(S^{'})
\end{eqnarray}
After this refinement, we get the sinogram domain output $\hat{S}$ for loss computing and next stage reconstruction.

\subsection{Sinogram Inner-Structure Loss}

% \begin{figure*}
% 	\begin{center}
% 		%\fbox{\rule{0pt}{2in} \rule{1\linewidth}{0pt}}
% 		\includegraphics[width=1\linewidth]{conjugate.pdf}
		
% 	\end{center}
% 	\caption{Illustration of conjugate sampling pairs in sinogram.}
% 	\label{conj_i}
% \end{figure*}
%Due to the special structure of sinogram, train the sinogram transformer only with traditional loss functions e.g., L1 loss and MSE loss, can be misleading. Besides, the inner-structure characteristics can very helpful while denoising.Therefore, we propose the sinogram inner-structure loss in this section for improving the denoising quality both in sinogram and image domain.

The SISL is designed in two levels: the global inner-structure loss $L_{C}$ and local inner-structure loss $L_{S}$:
\begin{eqnarray}
SISL=L_{C}+L_{S}
\end{eqnarray}

The global inner-structure utilizes conjugate sampling pairs in CT scans. This relation constrains the corresponding points to obtain the same value. Since this correlation exists in the whole sinogram, it can be used for maintaining the global inner-structure. The local inner-structure uses the second-order derivatives sparsity of sinogram. This feature indicates the correlation between the adjacent points in the sinogram which helps to maintain the local inner-structure. The details of each loss are shown as follows:

\subsubsection{Global Inner-Structure Loss} As shown in Fig.\ref{motivation} (a), conjugate sampling pairs receive X-rays with the same path so that their value should be the same. Let the sample of interest forms a detector angle $\gamma$ in view angle $\beta$. When the CT gantry is rotated to the view angle of ${\beta+\pi+2\gamma}$, the original sample of interest obtain the same value as the detector that forms an angle $-\gamma$. In the sinogram, it can be formulated as:

\begin{figure*}[!h]
	\begin{center}
		%\fbox{\rule{0pt}{2in} \rule{1\linewidth}{0pt}}
		\includegraphics[width=1\linewidth]{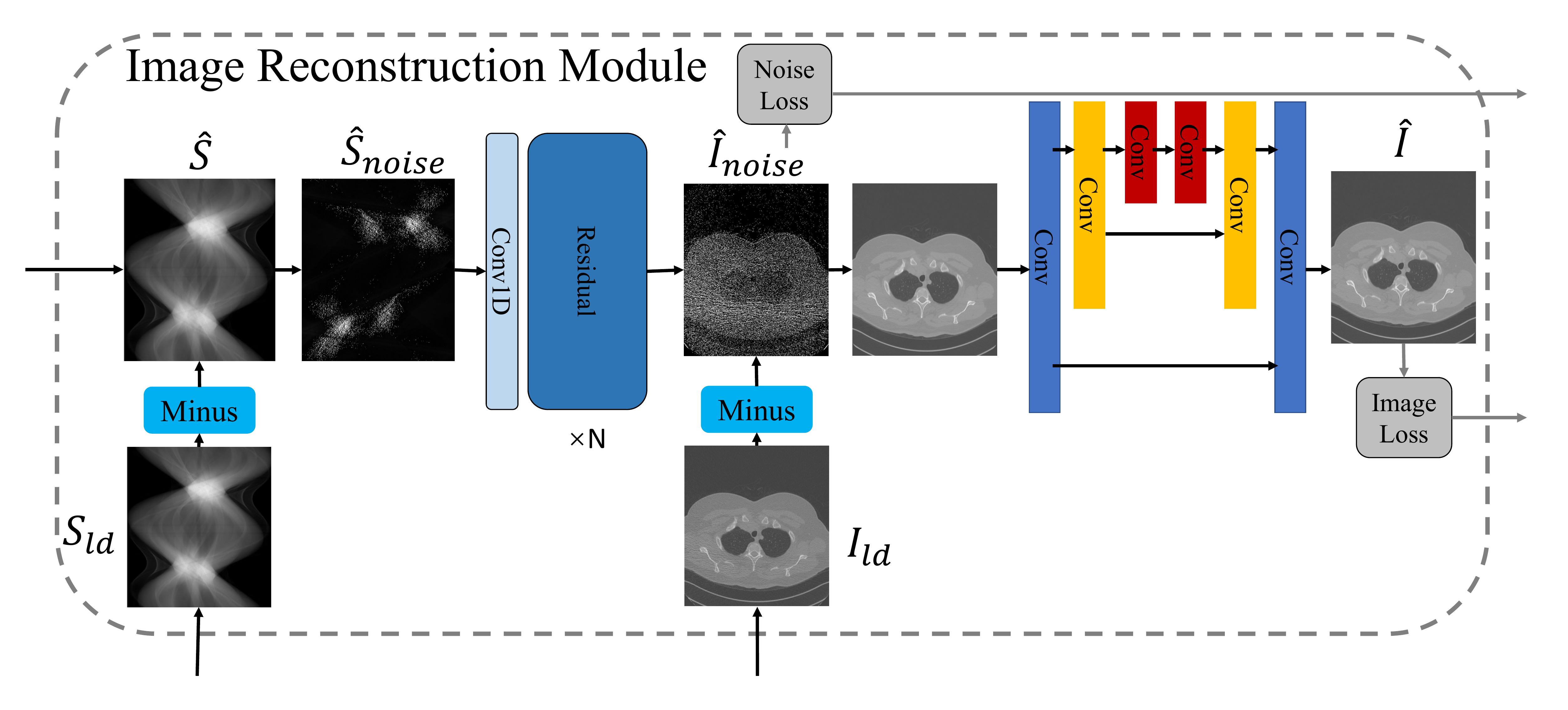}
		
	\end{center}
	\caption{The architecture of the image reconstruction module. The sinogram domain noise is converted into the image domain. This module use $S_{ld}$ minus $\hat{S}$ to generate the sinogram noise $\hat{S}_{noise}$ and transfer it into image domain $\hat{I}_{noise}$. Using $I_{ld}$ minus $\hat{I}_{noise}$, we can get the coarse denoised image. After the final image refinement, we can get the denoised image $\hat{I}$. }
	\label{imre}
\end{figure*}

\begin{eqnarray}
\label{conj}
S(\beta,\gamma)=S(\beta+\pi+2\gamma,-\gamma)
\end{eqnarray}
where $S$ is the sinogram. Let $P$ be the total number of projections in a $2\pi$ scan and $\gamma_{d}$ be the detector angle of one detector. Then Eq. \ref{conj} can be discretely rewritten as

\begin{eqnarray}
\label{conjd}
S(i,j)=S(i+int(\frac{(\pi+2j\gamma_{d})P}{2\pi}),-j)
\end{eqnarray}

 Thus, for each point $S(i,j)$ in $S$, its conjugate sampling should obtain the same value. Based on this fact, we can define the conjugate sinogram $\hat{S_{C}}$ of $\hat{S}$ using Eq. \ref{conjd}. Since the noise appears randomly in the imaging procedure, the conjugate relation would be broken in low-dose scans. As a consequence, the conjugate sinogram $\hat{S_{C}}$ would be different from $\hat{S}$ in some degree (depends on the noise level). By formulating a loss function based on the distance between $\hat{S_{C}}$ and $\hat{S}$, we can maintain the inner-structure and restrain noise while training. The global inner-structure loss can be formulated as:
\begin{eqnarray}
\label{conjr}
L_{C}=||\hat{S}-\hat{S}_{C}||
\end{eqnarray}

This loss calculates the Euclidean distance between $\hat{S_{C}}$ and $\hat{S}$. By minimizing this loss while network training, the noise can be effectively reduced. 
%\subsection{Sinogram Local Inner-structure Loss}
\subsubsection{Local Inner-Structure Loss}  Fig.\ref{motivation} (b) gives the visualization of horizontal second-order derivatives of the sinogram. As illustrated in \cite{xie2017robust}, the sinogram can be described as a piecewise linear configuration so that we can obtain that second-order derivatives of sinogram will be very sparse. To utilize this inner-structure in our deep learning framework, one feasible way is to use the Hessian penalty \cite{boyd2004convex,sun2015iterative} to introduce the sparsity of second-order derivatives. However, if the second-order derivative is over-sparsified while training, it would result in over-smooth in images. Since it's hard to find an appropriate hyper-parameter to control the sparsity, we rather use the loss between the low-dose sinogram and ground truth (\textit{i.e}, normal-dose sinogram). So, we define the local inner-structure loss of sinogram based on second-order derivatives:

\begin{eqnarray}
\label{Second-Order}
\begin{aligned}
&L_{S}=\sum_{i,j}{\sqrt{D_{ii}^{2}+D_{jj}^{2}+D_{ij}^{2}+D_{ji}^{2}}} , \\
&D_{ii}\triangleq \frac{\partial^2 \hat{S}(i,j)}{\partial i^2}-\frac{\partial^2 S_{nd}(i,j)}{\partial i^2},\\
&D_{jj}\triangleq \frac{\partial^2 \hat{S}(i,j)}{\partial j^2}-\frac{\partial^2 S_{nd}(i,j)}{\partial j^2},\\
&D_{ij}\triangleq \frac{\partial^2 \hat{S}(i,j)}{\partial i\partial j}-\frac{\partial^2 S_{nd}(i,j)}{\partial i\partial j},\\
&D_{ji}\triangleq \frac{\partial^2 \hat{S}(i,j)}{\partial j\partial i}-\frac{\partial^2 S_{nd}(i,j)}{\partial j\partial i}
\end{aligned}
\end{eqnarray}
where $D_{ii},D_{jj},D_{ij},D_{ji}$ are differences of second-order derivatives between output $\hat{S}$ and ground truth $S_{nd}$.

\begin{figure}
	\begin{center}
		\includegraphics[width=0.8\linewidth]{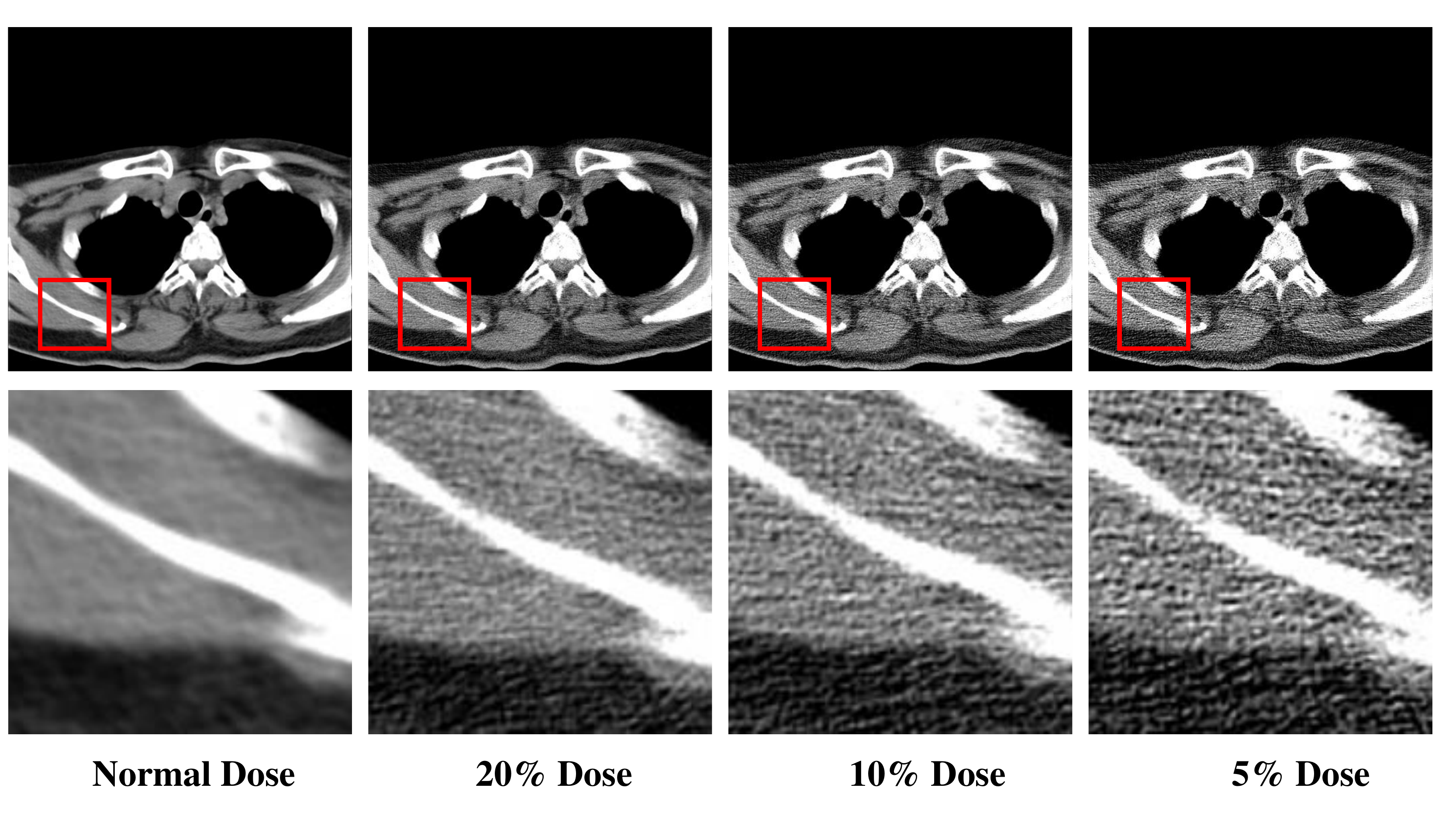}
	\end{center}
	\caption{Examples of LDCT images in Simulated Dataset under different dose levels. The LDCT images are generated using the noise inserting method in \cite{yu2012development}.}
	\label{dose_show}
\end{figure}

\begin{figure*}[!h]
	\begin{center}
		%\fbox{\rule{0pt}{2in} \rule{1\linewidth}{0pt}}
		\includegraphics[width=1\linewidth]{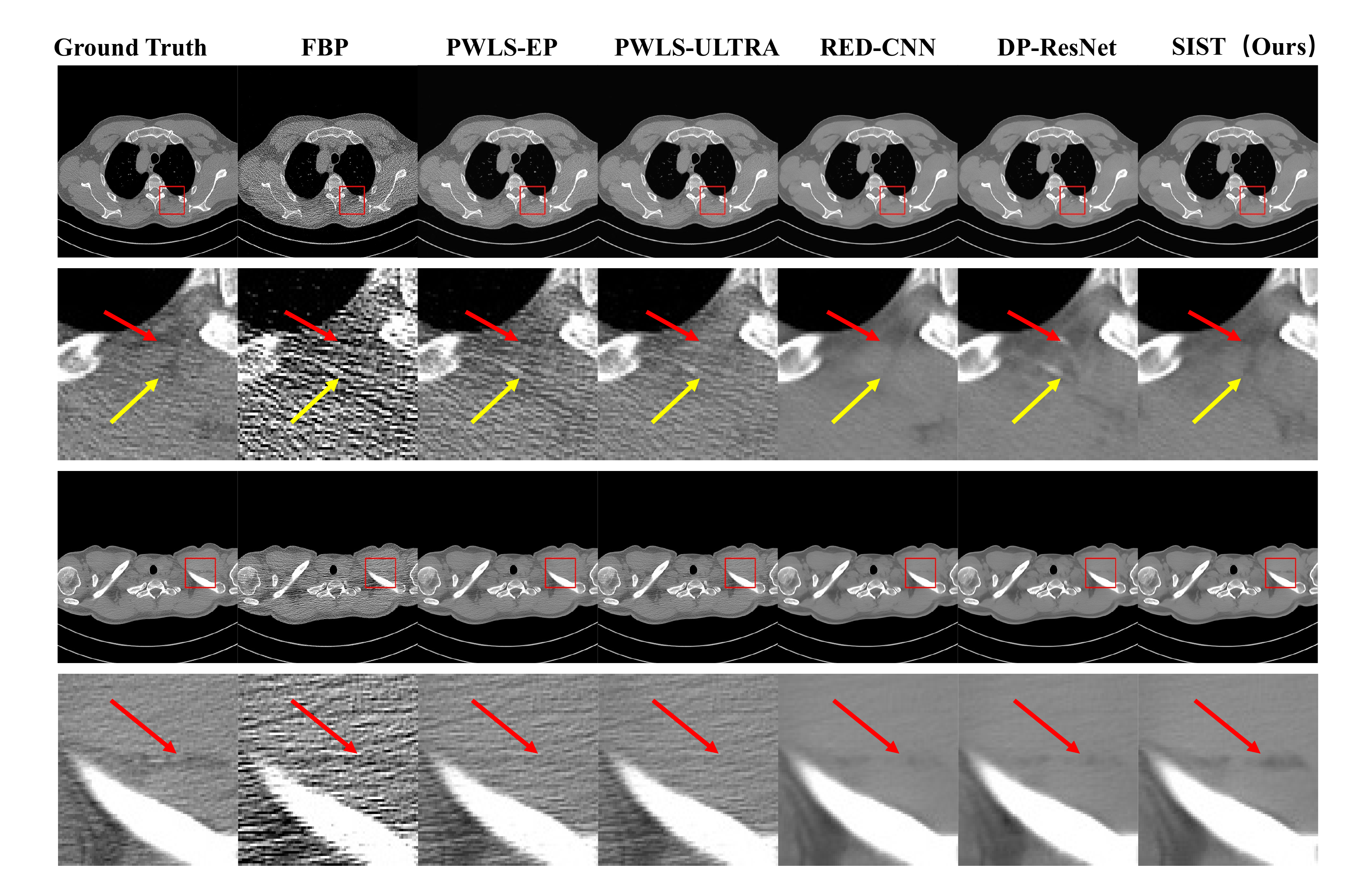}
		
	\end{center}
	\caption{Visual results in LDCT Dataset. The first and third rows are denoised images. The second and fourth rows are zoomed regions in red boxes. All images displayed are from the same window of [-460, 540]. }
	\label{compare}
\end{figure*}

\subsection{Image Reconstruction Module}
%In this part, we want to employ a deep learning based image reconstruction module so that we can back propagate the image loss to guide the sinogram training. 
The architecture above achieves effective sinogram domain denoising. To further apply image domain denoising complementarily, we propose the image reconstruction module. 
As shown in Fig.\ref{imre}, the sinogram domain noise is converted into the image domain. By removing the estimated noise from the LDCT image, a coarse denoised image is generated as the input of the later image denoising network. 

Since Anirudh \emph{et al.} \cite{anirudh2018lose} has successfully achieved domain transferring in a limited view CT restoring, we use a similar structure CNN as \cite{anirudh2018lose} but removing the sinogram complementing module.    
Instead of directly reconstructing the sinogram to the image, we first convert the sinogram noise into the image domain. Then we remove the estimated noise from the LDCT image and apply the image domain denoising. By using this design, the LDCT images can be input into the denoising network to add extra image domain information and guide the domain transfer. The learned reconstruction is actually between the sinogram noise and image noise:
\begin{eqnarray}
\hat{I}_{noise}=Residual(Conv1d(S_{ld}-\hat{S}))
\end{eqnarray}
where $\hat{I}_{noise}$ is the estimated noise between LDCT and NDCT images, $Conv1d$ is a 1D CNN embeds the sinogram noise into a latent space \cite{anirudh2018lose}. Then we use several residual blocks to transfer sinogram embeds into the image domain. finally, we remove the estimated image noise $\hat{I}_{noise}$ from LDCT images, and then input it into the next stage:
\begin{eqnarray}
\hat{I}=UNet(I_{ld}-\hat{I}_{noise})
\end{eqnarray}
where $\hat{I}$ is the final denoised image and $UNet$ is CNNs refer to \cite{ronneberger2015u}. This stage can further refine the image to achieve a better denoising performance.
 %Instead of directly reconstruct the sinogram using typical standard analytical reconstruct, e.g. FBP, we deploy the reconstruction into the deep learn framework. In the sinogram domain, traditional loss function cannot insure the reconstructed image quality because that a small difference in sinogram may become huge artifact after analytical reconstruction. 

%In stead of convert sinogram into reconstructed images, we choose to convert the sinogram noise into image noise:

\begin{figure*}[!h]
	\begin{center}
		%\fbox{\rule{0pt}{2in} \rule{1\linewidth}{0pt}}
		\includegraphics[width=1\linewidth]{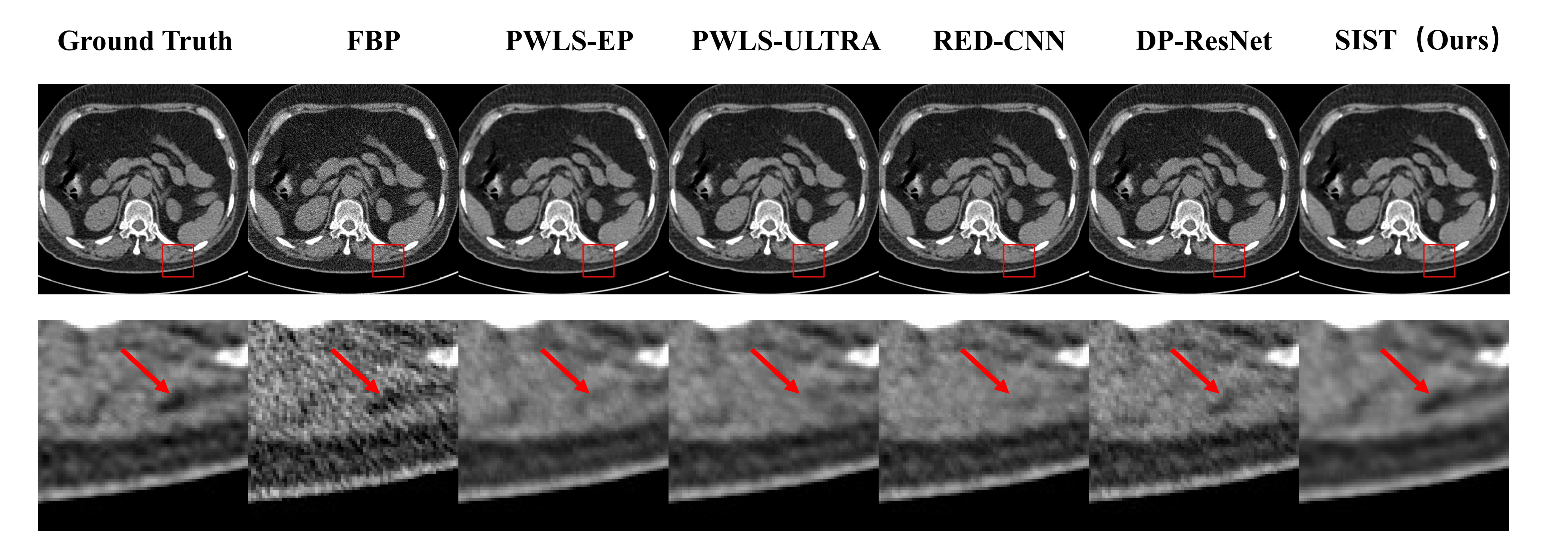}
		
	\end{center}
	\caption{Visual results of 20\% dose in Simulated Dataset. The first row is denoised images. The second row is zoomed regions in red boxes. All images displayed are from the same window of [-160, 240].  }
	\label{comp20}
\end{figure*}

\begin{figure*}[!h]
	\begin{center}
		%\fbox{\rule{0pt}{2in} \rule{1\linewidth}{0pt}}
		\includegraphics[width=1\linewidth]{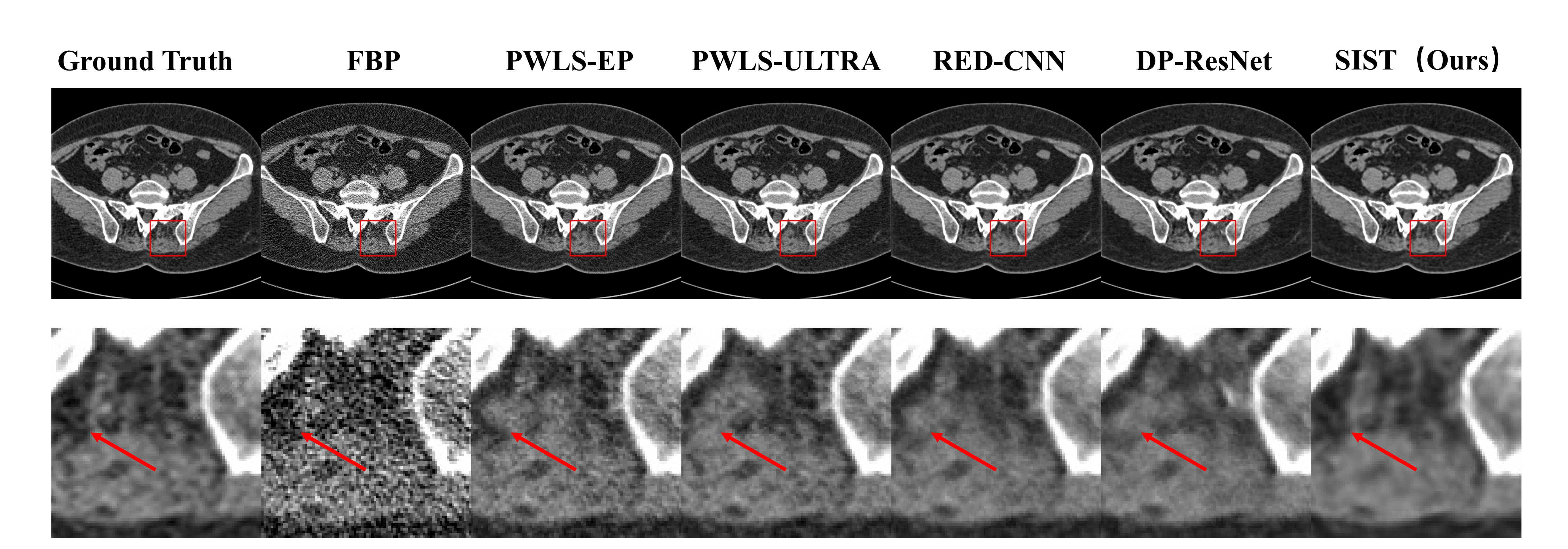}
		
	\end{center}
	\caption{Visual results of 10\% dose in Simulated Dataset. The first row is denoised images. The second row is zoomed regions in red boxes. All images displayed are from the same window of [-160, 240].  }
	\label{comp10}
\end{figure*}

\subsection{Total Loss Function}
In our proposed framework, The total loss function consists of four parts: sinogram denoising loss, sinogram inner-structure loss, image reconstruction loss, and image denoising loss. It can be formulated as:
\begin{eqnarray}
\label{loss}
\begin{aligned}
L=&||\hat{S}-S_{nd}||_{L1}+SISL+||\hat{I}-I_{nd}||_{L1}\\
&+||\hat{I}_{noise}-(I_{ld}-I_{nd})||_{L1}
\end{aligned}
\end{eqnarray}
 \\
In sinogram denoising, the L1 loss is used between the generated sinogram and ground-truth. Then, the SISL is added to maintain the inner-structure of the sinogram. In noise converting, we use the L1 loss between generated noise and the difference between LDCT and NDCT. For the image denoising, we also use the L1 loss.

\section{Experiments}
\subsection{Datasets}
We conduct experiments on two datasets to evaluate the performance of our proposed method:
\subsubsection{Low-Dose CT Image and Projection Dataset}
The \emph{Low-Dose CT Image and Projection Dataset}(LDCT Dataset) \cite{moen2021low} consists of CT patient scans from three common exam types: non-contrast head CT scans acquired for acute cognitive or motor deficit, low-dose non-contrast chest scans acquired to screen high-risk patients for pulmonary nodules, and contrast-enhanced CT scans of the abdomen acquired to look for metastatic liver lesions. It contains both the projection and image data from 150 clinically performed patient CT exams SOMATOM Definition CT system. the scanning parameters are as follows: 576 projection views evenly spanning a $360^{\circ}$ circular orbit were collected using a z-flying focal spot, resulting in 1152 projections within one rotation; the number of detector bins in each projection is 736. All 25141 samples have low-dose and normal-dose pairs. We randomly split 120 patients for training and the remains are for testing.

\subsubsection{Simulated Dataset}
The Simulated Dataset consists of CT images from spine CT exams. All 46 patients are collected from Qilu Hospital of Shandong University. Since only NDCT images are available in this dataset, we use the ASTRA tomography Toolbox\cite{van2016fast} to generate the projection data. In order to simulate the LDCT exams, the noise is inserted using a previously validated photon-counting model that incorporates the effect of the bowtie filter, automatic exposure control, and electronic noise \cite{yu2012development}. In this dataset, we generate the LDCT in different dose levels: 5\%, 10\%, and 20\% of the normal dose. The noise inserting function is shown as follows:
\begin{eqnarray}
P_{B}=P_{A}+\sqrt{\frac{1-a}{a}\frac{exp(P_{A})}{N_{0A}}(1+\frac{1+a}{a}\frac{N_{e}exp(P_{A})}{N_{0A}})\textbf{x}}
\end{eqnarray}
where $P_{A}$ is the logarithm-transformed projection data of the normal dose, $N_{0A}$ is the incident number of photons, $P_{B}$ is the simulated low dose projection data. $a$ indicates the dose levels of the simulated scans, $x$ is a random variable that follows a standard normal distribution, and $N_{e}$ is the noise-equivalent quanta of electronic noise. According to \cite{yu2012development},  $N_{0A}$ is set to $10^{5}$,  $N_{e}$ is set to 10. In our experiments, $a$ is set to $0.05$, $0.1$, and $0.2$ for dose-levels: 5\%, 10\%, and 20\% respectively. The training set contains 35 patients and the remains are for testing. 

\begin{table}[!h]

	\begin{center}
	    \linespread{1.5}
			\caption{ Experimental results for different methods on LDCT Dataset. Best results are \textbf{highlighted} and second results are \underline{underlined}. $\downarrow$ ($\uparrow$) means the lower (higher) the better. }
			\label{t1}
		\begin{tabular}{cccc}
			\hline
			Method & PSNR($\uparrow$)&SSIM($\uparrow$)&RMSE($\downarrow$) \\
			\hline
			FBP & 33.42\scriptsize{$\pm$3.78}&0.752\scriptsize{$\pm$0.078}&0.00640\scriptsize{$\pm$0.0077} \\
			PWLS-EP& 36.76\scriptsize{$\pm$1.45}&0.801\scriptsize{$\pm$0.026}&0.00436\scriptsize{$\pm$0.0021}\\
			PWLS-ULTRA& 38.98\scriptsize{$\pm$1.33}&0.902\scriptsize{$\pm$0.022}&0.00337\scriptsize{$\pm$0.0011}\\
			RED-CNN& 40.31\scriptsize{$\pm$1.43}&0.908\scriptsize{$\pm$0.013}&0.00289\scriptsize{$\pm$0.0013}\\
			
			DP-ResNet&\underline{40.92\scriptsize{$\pm$1.23}}&\underline{0.914\scriptsize{$\pm$0.011}}&\underline{0.00269\scriptsize{$\pm$0.0012}}\\
			%\hline
			SIST(Ours)&\textbf{41.80\scriptsize{$\pm$1.37}}&\textbf{0.916\scriptsize{$\pm$0.012}}&\textbf{0.00246\scriptsize{$\pm$0.0009}}\\
			%Ours & Makes one's heart Frob\\
			\hline
			
		\end{tabular}
	\end{center}
\end{table}

\begin{table*}[!h]
    
	\begin{center}
	
	        \linespread{1.5}
			\caption{ Experimental results for different methods on Simulated Dataset. Best results are \textbf{highlighted} and second results are \underline{underlined}. $\downarrow$($\uparrow$) means the lower (higher) the better. }
			\label{t2}
		%\resizebox{\textwidth}{15mm}{
		\scalebox{0.8}{
		\setlength{\tabcolsep}{1mm}{
		\begin{tabular}{cccccccccccc}
			\hline
			dose&\multicolumn{3}{c}{20\%}&& \multicolumn{3}{c}{10\%}&&\multicolumn{3}{c}{5\%}\\
            \cline{2-4} \cline{6-8} \cline{10-12}
			Method&PSNR($\uparrow$)&SSIM($\uparrow$)&RMSE($\downarrow$)&& PSNR($\uparrow$)&SSIM($\uparrow$)&RMSE($\downarrow$)&&
			PSNR($\uparrow$)&SSIM($\uparrow$)&RMSE($\downarrow$) \\
			\hline
			
			FBP&35.72\scriptsize{$\pm$5.17}&0.846\scriptsize{$\pm$0.096}&0.0189\scriptsize{$\pm$0.0082}&&31.78\scriptsize{$\pm$4.60}&0.743\scriptsize{$\pm$0.130}&0.0291\scriptsize{$\pm$0.0072}&&28.22\scriptsize{$\pm$3.87}&0.625\scriptsize{$\pm$0.150}&0.0422\scriptsize{$\pm$0.0067}\\

			PWLS-EP&38.88\scriptsize{$\pm$1.51}&0.901\scriptsize{$\pm$0.021}&0.0073\scriptsize{$\pm$0.0021}&
			&37.68\scriptsize{$\pm$1.47}&0.868\scriptsize{$\pm$0.014}&0.0083\scriptsize{$\pm$0.0018}&
			&36.88\scriptsize{$\pm$1.49}&0.843\scriptsize{$\pm$0.018}&0.0091\scriptsize{$\pm$0.0021}\\
			
			PWLS-ULTRA&40.28\scriptsize{$\pm$1.01}&0.942\scriptsize{$\pm$0.007}&0.0062\scriptsize{$\pm$0.0008}&
			&38.93\scriptsize{$\pm$1.02}&0.904\scriptsize{$\pm$0.0009}&0.0072\scriptsize{$\pm$0.0007}&
			&37.22\scriptsize{$\pm$1.01}&0.893\scriptsize{$\pm$0.0011}&0.0087\scriptsize{$\pm$0.0012}
			\\
			%d
			RED-CNN&42.72\scriptsize{$\pm$1.12}&\underline{0.970\scriptsize{$\pm$0.008}}&0.0047\scriptsize{$\pm$0.0008}&
			&\underline{42.31\scriptsize{$\pm$1.13}}&\underline{0.962\scriptsize{$\pm$0.012}}&\underline{0.0048\scriptsize{$\pm$0.0009}}&
			&40.67\scriptsize{$\pm$1.22}&0.948\scriptsize{$\pm$0.008}&0.0059\scriptsize{$\pm$0.0012}
			\\
			%d
			DP-ResNet&\underline{43.83\scriptsize{$\pm$1.17}}&0.960\scriptsize{$\pm$0.007}&\underline{0.0041\scriptsize{$\pm$0.0008}}&
			&42.03\scriptsize{$\pm$1.16}&0.953\scriptsize{$\pm$0.011}&0.0051\scriptsize{$\pm$0.0010}&
			&\underline{41.43\scriptsize{$\pm$1.65}}&\underline{0.952\scriptsize{$\pm$0.014}}&\underline{0.0054\scriptsize{$\pm$0.0015}}

			\\
			%\hline
			SIST(Ours)&\textbf{44.74\scriptsize{$\pm$1.12}}&\textbf{0.981\scriptsize{$\pm$0.006}}&\textbf{0.0037\scriptsize{$\pm$0.0007}} &  &\textbf{43.70\scriptsize{$\pm$1.12}}&\textbf{0.976\scriptsize{$\pm$0.008}}&\textbf{0.0042\scriptsize{$\pm$0.0006}}&
			&\textbf{42.62\scriptsize{$\pm$1.14}}&\textbf{0.973\scriptsize{$\pm$0.009}}&\textbf{0.0047\scriptsize{$\pm$0.0007}}
			\\
			%Ours & Makes one's heart Frob\\
			\hline
			
		\end{tabular}}}
	\end{center}
\end{table*}
\begin{figure*}[!h]
	\begin{center}
		%\fbox{\rule{0pt}{2in} \rule{1\linewidth}{0pt}}
 		\includegraphics[width=1\linewidth]{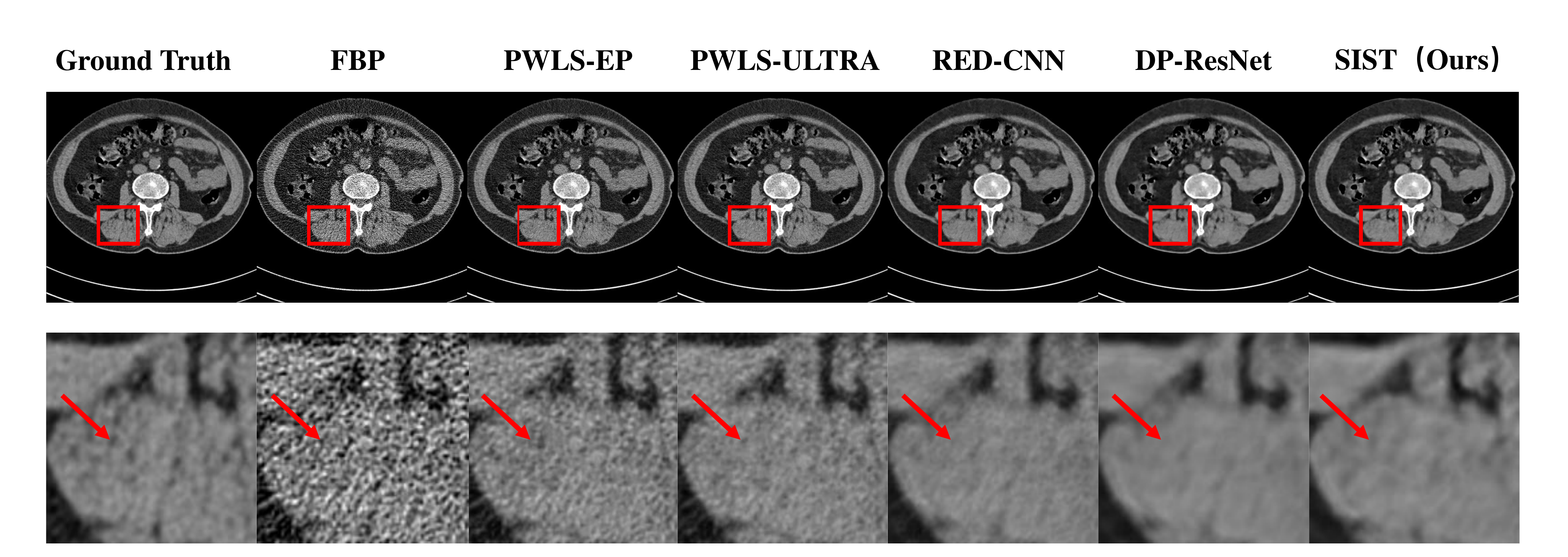}
		
	\end{center}
	\caption{Visual results of 5\% dose in Simulated Dataset. The first row is denoised images. The second row is zoomed regions in red boxes. All images displayed are from the same window of [-160, 240]. }
	\label{comp5}
\end{figure*}
\begin{table*}[!h]:
	\begin{center}
	    \linespread{1.5}
			\caption{ Experimental results for ablation study. The best results are \textbf{highlighted}. $\downarrow$ ($\uparrow$) means the lower (higher) the better. }
			\label{t3}
			
		\begin{tabular}{ccccccc}
			\hline
			Method &Sinogram&Global Loss&Local Loss& PSNR($\uparrow$)&SSIM($\uparrow$)&RMSE($\downarrow$) \\
			\hline
			Unet & &&&38.09\scriptsize{$\pm$1.44}&0.866\scriptsize{$\pm$0.022}&0.00391\scriptsize{$\pm$0.0019} \\
			SIST& \checkmark&&& 40.16\scriptsize{$\pm$1.39}&0.903\scriptsize{$\pm$0.018}&0.00292\scriptsize{$\pm$0.0011}\\
			SIST+$L_{C}$&\checkmark &\checkmark&&41.27\scriptsize{$\pm$1.41}&0.912\scriptsize{$\pm$0.013}&0.00254\scriptsize{$\pm$0.0011}\\
			SIST+$L_{C}$+$L_{S}$&\checkmark &\checkmark&\checkmark&\textbf{41.80\scriptsize{$\pm$1.37}}&\textbf{0.916\scriptsize{$\pm$0.012}}&\textbf{0.00246\scriptsize{$\pm$0.0009}}\\
			
			%MAP-NN& &&&\underline{40.33}&0.913&\underline{0.00278}\\
			%\hline
			%Ours & &&&\textbf{41.80}&\textbf{0.916}&\textbf{0.00246}\\
			%Ours & Makes one's heart Frob\\
			\hline
			
		\end{tabular}
	\end{center}
\end{table*}

\begin{figure*}[!h]
\begin{center}
	\includegraphics[width=0.75\linewidth]{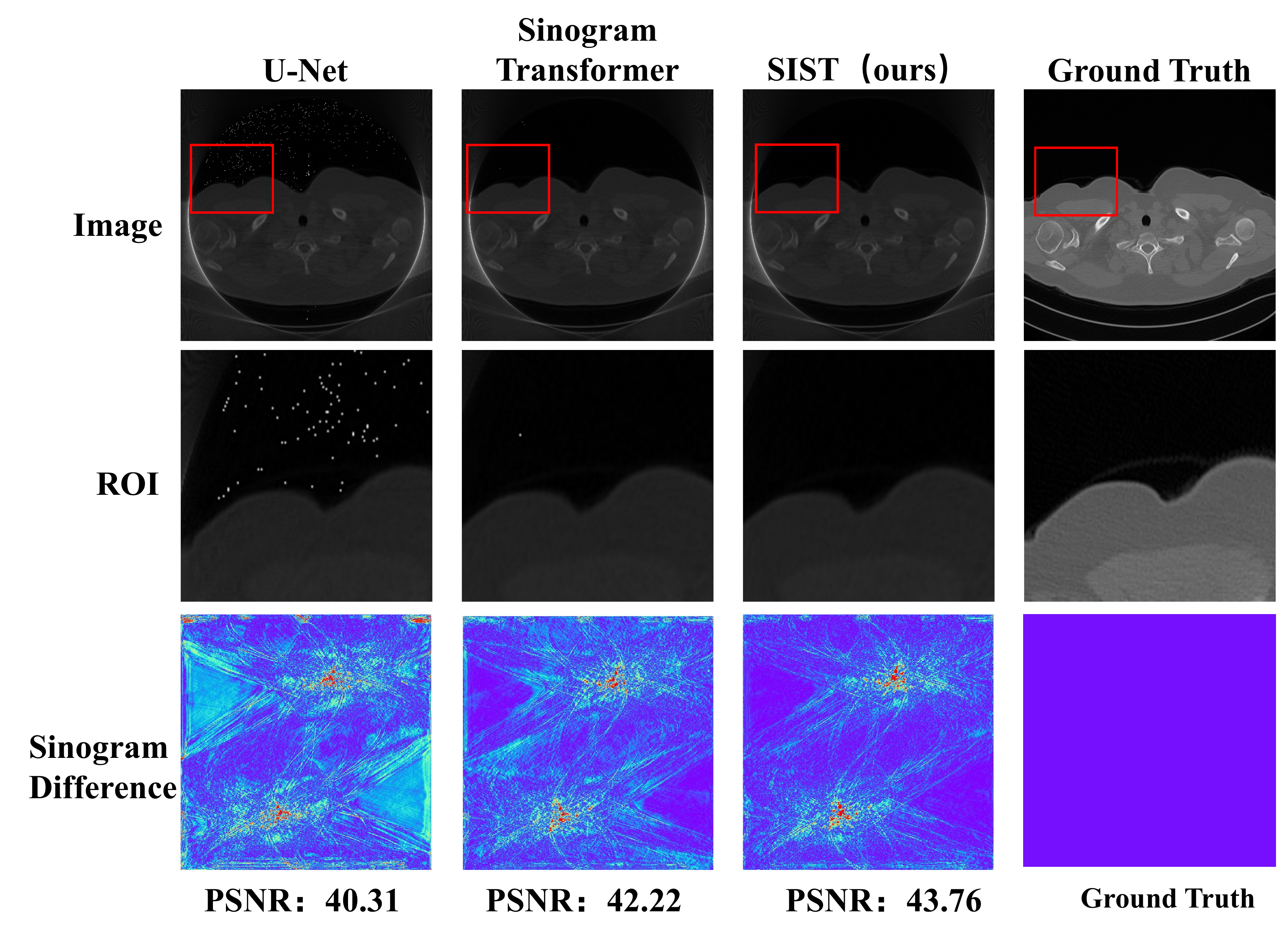}
	\caption{Visual results of ablation study that verify the effectiveness in sinogram domain. We compare the sinogram domain denoising performance with a CNN-based U-Net, a single sinogram transformer, and our proposed method which uses the sinogram transformer with inner-structure loss. We reconstruct images of all three methods using FBP. The first row shows the reconstructed images. ROIs in the red box are zoomed in the second row. The third row shows the sinogram difference with ground truth (the bluer, the better). }
	\label{ablation}
	\end{center}
\end{figure*}

\subsection{Implementation Details}
In the sinogram transformer module, the head uses two linear layers and converts the sinogram into dimension 1024. In the tail, the residual block consists of two residual convolutional layers. The MSA in the transformer encoder has 6 heads and  6 layers. The network is implemented by PyTorch with python 3.6 on 4 NVIDIA RTX 2080Ti. We used ADAM with momentum for optimization and set the initial learning rate to be $10^{-5}$, the first-order momentum to be 0.9, and the second momentum to be 0.999. A stepLR is used with a 0.7 decay rate every 10 epoch. 

For image quality evaluation, three metrics are used in our experiments: peak signal to noise ratio (PSNR), structural similarity index measure (SSIM) and root mean square error (RMSE). 
%\subsubsection{Evaluation Metrics}
 %To objectively evaluate the LDCT denoising performance, we employ three metrics for evaluation: peak signal-to-noise ratio (PSNR),Structural Similarity (SSIM) and root-mean-square error (RMSE). 

\subsection{Comparison with Other Methods}
In order to verify the performance of the proposed method,
we conduct the comparison experiment with several state-of-the-art methods, i.e., PWLS-EP, PWLS-ULTRA, RED-CNN, and DP-ResNet. Both the PWLS-EP\cite{zheng2018pwls} and PWLS-ULTRA\cite{zheng2018pwls} are iterative reconstruction methods that utilize the low-dose sinogram. RED-CNN\cite{chen2017low} and DP-ResNet\cite{yin19} are CNN-based methods. RED-CNN only denoise in the image domain and DP-ResNet uses both sinograms and images. 
\subsubsection{LDCT Dataset}
We first conduct the experiments on the Low-Dose CT Image and Projection Dataset.
Tab.\ref{t1} gives quantitative results of all methods. Note that FBP in Tab.\ref{t1} are images directly reconstructed from a low-dose sinogram using FBP. From these results, we can observe that deep learning methods outperform iterative reconstruction methods thanks to the delicate designed CNN architectures. By further utilizing the sinogram inner-structure information, our method achieves the best performance in all three metrics. 
Visual results are shown in Fig.\ref{compare}. It can be observed that our proposed method outperforms others.

\subsubsection{Simulated Dataset}
To further evaluate the performance of the proposed method on different dose levels, we generate the projection data and insert noise follow \cite{yu2012development}. We evaluate the performance on three dose levels: 20\%, 10\%, and 5\%. The quantitative results are shown in Tab.\ref{t2}. From the table, we can observe that our methods achieve the best results in all dose levels: 20\%, 10\%, and 5\%. Same as the LDCT dataset, the deep learning methods show superior performance that iterative reconstruction methods. Moreover, our method still has great performance at the very low dose situation (5\% dose), which even over the path the second method in 10\% dose. Fig.\ref{comp20}, Fig.\ref{comp10}, and  Fig.\ref{comp5} show the visual results for dose levels 20\%, 10\%, and 5\%. From these visual results, we can observe that our method can preserve more details while denoising.

\subsection{Ablation Study}

To further explore the effectiveness of different components, we conduct the ablation study in this section.

\subsubsection{Effectiveness in Image Domain}  Compare to typical deep learning methods, the proposed framework focuses on the inner-structure of the sinogram to improve the image domain denoising quality. So, one important question is that \emph{how much does the inner-structure impact the image quality? } Tab.\ref{t3} shows results of using different components. We first remove all sinogram domain components and use the rest part (U-Net) to train and test only in the image domain. Without any specific design, the PSNR achieves 38.09 $dB$ of pure U-Net CNNs. After we add the sinogram denoising transformer and image reconstruction module, the performance improved significantly in all three metrics. To further utilize the inner-structure of the sinogram, the global inner-structure loss is added to utilize conjugate projection pairs in the sinogram. This loss helps improve the performance as shown in Tab.\ref{t3}. At last, we add the local inner-structure loss to maintain the second-order sparsity of the sinogram, and the performance is further improved.
\subsubsection{Effectiveness in Sinogram Domain} In this part, we verify the effectiveness of our method in improving the sinogram denoising. For comparison, we use the U-Net and sinogram transformer module to train with low-dose/normal-dose sinogram pairs. Since we care more about the quality of reconstructed images, we evaluate sinograms by applying FBP reconstruction. Note that our proposed method use end-to-end training to directly get the reconstructed image. For a fair comparison, we still apply FBP on the intermediate sinogram output.  Fig.\ref{ablation} shows the example of the results. As we can observe, even though all the denoised sinograms are of high similarity to the ground truth, there still are obvious artifacts in reconstructed images. Compare to the CNN-based U-Net, the sinogram transformer can better extract the structure information in the sinogram and improve the quality. By further adding the inner-structure loss of sinogram, artifacts in images are considerably reduced.

\section{Conclusion}
In this paper, we introduce the sinogram inner-structure transformer which utilizes the sinogram domain information to assist the LDCT denoising in the image domain. In order to improve the sinogram quality while training, we introduce the inner-structure loss to maintain the special structure of the sinogram. The inner-structure loss considers both the global and local inner-structure. Thus, the proposed SIST achieves a great performance in the LDCT denoising task. Experiments on two datasets also show that our method can significantly improve the image quality. Ablation studies verify the effectiveness of the components in our proposed framework. 
In the future, we will explore more inner-structure in the sinogram for further improvement on the LDCT.

%Bibliography
\bibliographystyle{unsrt}  
\bibliography{references}

\end{document}